\documentclass[10,pt,prl,twocolumn,showpacs]{revtex4}

\usepackage{graphicx}
\usepackage{dcolumn}
\usepackage{amsmath}
\usepackage{epsfig}
\usepackage{multirow}

\begin{document}

\title{Analysis of Rotation Curves in the framework of the Gravitational Suppression model}
\author{Christiane Frigerio Martins and Paolo Salucci}
\email{martins,salucci@sissa.it}
\affiliation{SISSA, Via Beirut 4, 34014 Trieste, Italy}

\begin{abstract}
We present an analysis of suitable rotation curves (RCs) of eight galaxies, aimed at checking the consistency and universality of the gravitational suppression (GraS) hypothesis, a phenomenological model for a new interaction between dark matter and baryons.
Motivated by the puzzle of the core versus cusp distribution of dark matter in the center of halos, this hypothesis claims to reconcile the predictions from N-body $\Lambda$ cold dark matter simulations with kinematic observations.
The GraS model improves the kinematic fitting residuals, but the mass parameters are unphysical and put the theory in difficulty.  
\end{abstract}

\pacs{95.35.+d, 04.50.+h, 98.62.Gq}
\maketitle

The gravitational suppression hypothesis \cite{PM} is a phenomenological model that addresses the complex understanding of the dark matter distribution on small, subgalactic scales.
High-resolution radio observations from spiral galaxies, along with their optical rotation curves (RCs), suggest that the dark matter is distributed in spherical halos with nearly constant density cores (see, e.g., \cite{gentile,salucci,donato} and references therein).
On the other hand, theoretical predictions from the well-known  N-body $\Lambda$CDM simulations (e.g., \cite{nfw}) present a steep density distribution profile in the centre of the halos: 
\begin{equation}
\rho_{halo}(r)=\frac{\rho_{s}}{(r/r_{s})(1+r/r_{s})^{2}}.
\end{equation}
$r_{s}$ is a scale radius and $\rho _{s}$ its characteristic density, in principle independent, but found related within a reasonable scatter through the halo mass, by the Bullock \emph{et al.} \cite{bullock} equation: $c\equiv R_{vir}/r_{s}\sim18(\frac{M_{vir}}{10^{11}M_{\odot}})^{-0.13}$,
where $c$ is a concentration parameter and $R_{vir}$ and $M_{vir}$ are the virial radius and mass.
Mass models with a Navarro-Frank-White (NFW) density profile, given in Eq.(1), have two serious kinds of difficulty in reproducing the observed RCs: 
a) the fit is not satisfactory, i.e., $\chi^{2}_{red}\gg 1$ (see, e.g., \cite{gentile} and references therein); b) the values of the parameters of the best-fit mass models are clearly unphysical.
In detail, the values for the halo mass result much higher than those we obtain from weak lensing halo models \cite{wlensing} and from the analysis of galaxy baryonic mass function \cite{shankar}: $M_{halo}\approx 3\times 10^{12}M_{\odot}\: (L_{B}/10^{11}L_{\odot})^{1/2}$. 
In the same way the values of the disk mass-to-light ratio result much lower than those derived from colours of spirals \cite{shankar,bruzual,luminosity,LB}: $\log\: (M_{D}/M_{\odot})\approx -1.6+1.2 \log \:(L_{B}/L_{\odot}) $, i.e., $0.7< M_{D}/L_{B} < 4$.

Several solutions have been proposed for the above issue, most of them related either to a better comprehension of structure formation (e.g., \cite{chiara}) or to new fundamental physics (e.g., \cite{spergel}).
Alternatively, the presence of noncircular motions in galaxies has been advocated to reconcile (up tp $70\%$ in the Low Surface Brightness of) the observed kinematics with the cuspy density profile (e.g., \cite{hayashi04,hayashi06}, but see also \cite{47}). 

The original proposal by Piazza and Marinoni (PM) GraS model, instead,  modifies the usual Newtonian potential of the dark matter felt by baryonic test particles in such a way that the NFW kinematics and the observed one  become in agreement.
According to PM, the NFW profile is used because GraS does not affect the dark matter dynamics, but only the dynamics in the mixed sector dark matter-baryons, so both primordial dark matter perturbations and halo formation are unaffected, and well-known N-body simulation results can be assumed.
The idea is adding a Yukawa contribution to the gravitational potential
\begin{equation}
\nabla^{2}\phi_{Newton}=4\pi G\: (\rho_{baryons}+\rho_{halo}),
\end{equation}
from a hypothetical short-range interaction just between dark and luminous matter
\begin{equation}
(\nabla^{2}-\lambda^{-2})\:\phi_{Yukawa}=4\pi G\:\rho_{halo},
\end{equation}  
where $\lambda$ is a \textit{scale range} parameter. The effect is damping the gravitational interaction on small scales.
The final potential is then
\begin{equation}
\phi_{halo}=\phi_{Newton}+\alpha \: \phi_{Yukawa}.
\end{equation}
$\alpha$ is a \textit{strength} parameter and taken to be $-1$ in order to have the maximum possible gravitational suppression \cite{veneziano}.
The circular velocity  is related to the potential by $V_{halo}^2=V_{halo,\: Newton}^2+V_{halo,\: Yukawa}^2=r \: |d\phi_{halo}/dr|$.

In  PM model, for  a (small) sample of RCs  of  Low Surface Brightness galaxies GraS was able to eliminate the above core versus cusp discrepancy.
However, in order to allow a   simple analytic calculation, they have  taken   a number of assumptions and  approximations.  In detail,  the contribution to the gravitational potential from baryons (stars and HI disk)  was neglected and the dark matter distribution was modeled with the  simple form $\rho_{halo}(x)=\rho_{0}x^{-\beta}$, rather than by Eq. (1). 
Further support to GraS was given in \cite{piazza} where the dispersion velocity of two spheroidal dwarfs (Fornax and Draco) were studied in this scenario.
However, both large errors in the kinematic measurements and large geometric and orbital  uncertainties of the employed  mass model, limited the relevance of their findings. 
\newpage
\begin{widetext}
\begin{centering}
\begin{table*}
\caption{Parameters of the mass models.}
\begin{tabular}{lcccccccrc}\hline\hline
Galaxy&$L_{B}\:(L_{\odot}$)&Mass model&$M_{D}/L_{B}$&$M_ {halo}\:(M_{\odot})$&$\chi^{2}_{red}$&$r_{s}\:(kpc)$&$\rho_{s}\:(10^{4}\rho_{crit,\:0})$&$M_{D}\:(M_{\odot})$&\emph{c}\\\hline

\multicolumn{5}{l}{Positive results}\\

\multirow{2}{*}{ESO 116-G12}&\multirow{2}{*}{$4.6\times 10^{9}$}&NFW&\textbf{0.1}&3.8$\times 10^{11}$&\textbf{2.8}&$14.5\pm14$&$4.0\pm6.6$&$(4.2\pm27)\times 10^{8}$&13\\
&&NFW+GraS&0.3&1.5$\times 10^{11}$&1&$5.1\pm2.3$&$26\pm25$&$(1\pm1.7)\times 10^{9}$&26\\

\multirow{2}{*}{UGC 10981}& \multirow{2}{*}{$1.2\times 10^{11}$}&NFW&1.5&2.6$\times 10^{11}$&\textbf{4.2}&$8\pm2.9$&$13\pm9$&$(1.8\pm0.3)\times 10^{11}$&21\\
&&NFW+GraS&0.4&7.7$\times 10^{11}$&2.5&$4.2\pm0.3$&$180\pm40$&$(4.9\pm4.4)\times 10^{10}$&55\\

\multicolumn{5}{l}{Negative results: unphysical parameters}\\
\multirow{2}{*}{DDO 47}&\multirow{2}{*}{$10^{8}$}&NFW&$<0.2$&$\mathbf{7.4\times 10^{12}}$&1.9&$176\pm10$&$0.12\pm0.1$&$<2.3\times10^{7}$&2.8\\
&&NFW+GraS&0.5&$\mathbf{8.1\times 10^{11}}$&0.4&$26\pm18$&$1.8\pm1.4$&$(4.5\pm2.2)\times 10^{7}$&9.2\\

\multirow{2}{*}{NGC 6822}&\multirow{2}{*}{$1.6\times 10^{8}$}&NFW&$\mathbf{<0.04}$&1.7$\times 10^{12}$&2.3&$87\pm49$&$0.19\pm0.12$&$<6.7\times10^{6}$&3.5\\
&&NFW+GraS&$\mathbf{<0.02}$&2.5$\times 10^{10}$&0.5&2.9$\pm0.1$&24$\pm0.7$&$<2.9\times10^{6}$&26\\

\multirow{2}{*}{ESO 79-G14}&\multirow{2}{*}{$2\times 10^{10}$}&NFW&\textbf{0.3}&$\mathbf{3.9\times 10^{13}}$&\textbf{5}&$330\pm1400$&$0.1\pm0.49$&$(6.4\pm1.9)\times 10^{9}$&2.6\\
&&NFW+GraS&$\mathbf{0.3}$&$1.1\times 10^{12}$&2&$22.9\pm6$&$3.2\pm1.4$&$(6\pm0.9)\times 10^{9}$&11.2\\

\multirow{2}{*}{UGC 8017}&\multirow{2}{*}{$4\times 10^{10}$}&NFW&1&$\mathbf{4.4\times10^{17}}$&\textbf{4}&$379\pm3600$&$150\pm60$&$(3.8\pm0.8)\times 10^{10}$&51\\
&&NFW+GraS&1.1&$\mathbf{1.5\times10^{14}}$&1.6&$22\pm9$&$250\pm50$&$(4.4\pm0.3)\times 10^{10}$&62\\

\multirow{2}{*}{UGC 11455}&\multirow{2}{*}{$4.5\times 10^{10}$}&NFW&1.4&3.6$\mathbf{\times 10^{13}}$&\textbf{7.2}&$121\pm13$&$0.9\pm0.1$&$(7\pm2)\times 10^{10}$&7\\
&&NFW+GraS&$\mathbf{<0.2}$&3.2$\times 10^{12}$&3.9&$13.7\pm0.5$&$28\pm2.6$&$<10^{10}$&27\\

\multicolumn{5}{l}{Negative result: no change}\\
\multirow{2}{*}{M 31}&\multirow{2}{*}{$2\times 10^{10}$}&NFW&6.5&1.4$\times 10^{12}$&2&$28.5\pm1$&$2.2\pm0.1$&$(1.3\pm0.1)\times 10^{11}$&10\\
&&NFW+GraS&7&1.4$\times 10^{12}$&2.2&$31\pm1.1$&$1.8\pm0.1$&$(1.4\pm0.1)\times 10^{11}$&9.2\\\hline\hline
\end{tabular}
\end{table*}
\end{centering}
\end{widetext}

In the present analysis of GraS we abandon the above approximations and test a wider and fairer sample of spirals.
An in-depth review of the GraS model is beyond the scope of this Letter.
Our goal is to perform a check of GraS.
First, we assume the exact  NFW profile. Second, we consider the baryonic contribution, so that the total potential is
\begin{equation}
\phi_{model}=\phi_{halo}+\phi_{disk}+\phi_{gas},
\end{equation}
where the sum of the last two terms is $\phi_{baryons}$.  
This leads to $V_{model}^2=V_{halo}^2+V^2_{disk}+V^2_{gas}$.
Finally, we use a sample of high-resolution RCs of Low and High Surface Brightness galaxies, in order to investigate the consistency and universality of the model.

Our sample represents the best available RCs to study the mass distribution of dark matter and it has been used in works concerning the core versus cusp discrepancy controversy \cite{gentile, corbelli}.
The sample includes nearby Low and High Surface Brightness galaxies, all poorly fitted by mass models with NFW halos that also have unphysical values for their best-fit mass parameters: DDO 47 \cite{47}; ESO 116-G12, ESO 79-G14 \cite{gentile}; NGC 6822 \cite{weldrake}; UGC 8017, UGC 10981, UGC 11455 \cite{vogt}; M 31 \cite{corbelli}.
Let us notice that in some cases H$_\alpha$ and HI RCs are  both available and they  agree well where they coexist.
Moreover the RCs we analyse are smooth, symmetric and extended to large radii. 
\begin{centering}
\begin{figure}[htbp]
\includegraphics[width=8.6cm]{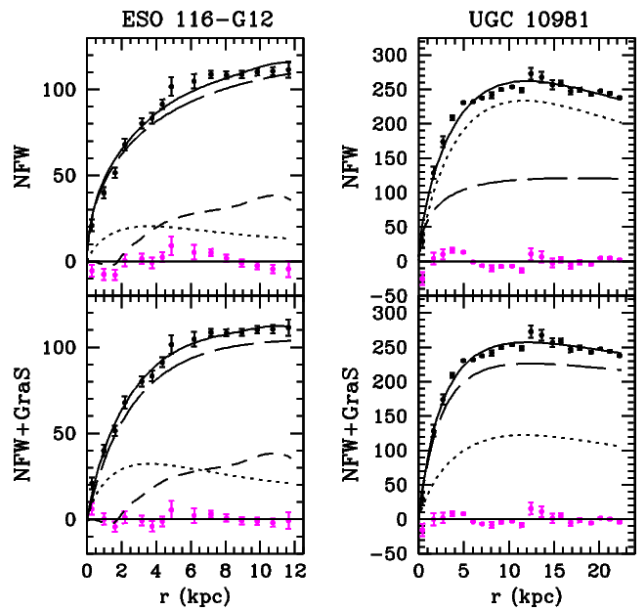}
\vskip  -0.25truecm
\caption{Galaxies in which GraS eliminates the core versus cusp discrepancy controversy. $Y$ axis is the velocity in km/s. The solid line represents the best-fit mass model, the long-dashed line is the contribution of the dark matter halo, and the dotted and short-dashed lines are those of the stellar and gaseous disks. Below the  RCs, we plot the residuals ($V_{obs}-V_{model}$).}
\label{fig:one}
\end{figure} 
\end{centering}

We decompose the total circular velocity into stellar, gaseous and halo contributions, according to Eqs.($1$)-($5$), where the latter contains the additional dark matter-baryons interaction.
Available photometry shows that the stars in our sample of galaxies are distributed in a thin disk, with exponential surface density profile $\Sigma_{D}(r)=(M_{D}/2 \pi R_{D}^{2})\: e^{-r/R_{D}}$, where $M_{D}$ is the disk mass and $R_{D}$ is the scale length.
The circular velocity contribution is given by $V_{disk}^{2}(r)=(G M_{D}/2R_{D}) \:  x^{2}B(x/2)$, where $x\equiv r/R_{D}$ and $G$ is the gravitational constant.
The quantity $B=I_{0}K_{0}-I_{1}I_{1}$ is a combination of Bessel functions \cite{freeman}.
The contribution of the gaseous disk is directly derived from the HI surface density distribution.

In a first step, the RCs are $\chi^{2}$ best-fitted with the following free parameters: disk mass, NFW scale radius and characteristic density, and scale range of GraS. 
Then we redo the analysis fixing the GraS scale range parameter at the mean value found  of $\lambda=3.1$\:kpc.
Notice that the published mean value of PM for $\lambda$ is quite different from ours as an effect of their simplifications: $\lambda=1.1$\:kpc. 
Our value is the most favourable for the PM model: different values of $\lambda$ leads to worse performance.

The test goes against the  GraS model.
For the RCs of our sample the NFW mass halo model fails to reproduce data according to the usual pattern explained in the introduction.
Data, not surprisingly, points to dark matter halos having inner density cores.
Applying a Yukawa potential to the cuspy NFW halo does not solve this discrepancy.
The cusp is erased and RCs are fitted very well, but this success is illusory in that the corresponding values of the parameters of the best-fit mass model remain unphysical.
In table I we show the results of the test. We give: the values of the parameters of the mass model and global properties of the galaxies. 
$\chi^{2}_{red}$ is calculated with average typical velocity errors. In bold, unphysical values for halo mass and mass-to-light ratio, and $\chi^{2}_{red}>2.5$.  
The critical density of the Universe today is taken to be $\rho_{crit,\:0}=10^{-29}g/cm^{3}$.
\begin{widetext}
\begin{centering}
\begin{figure*}[htbp]
\includegraphics[width=14.85cm]{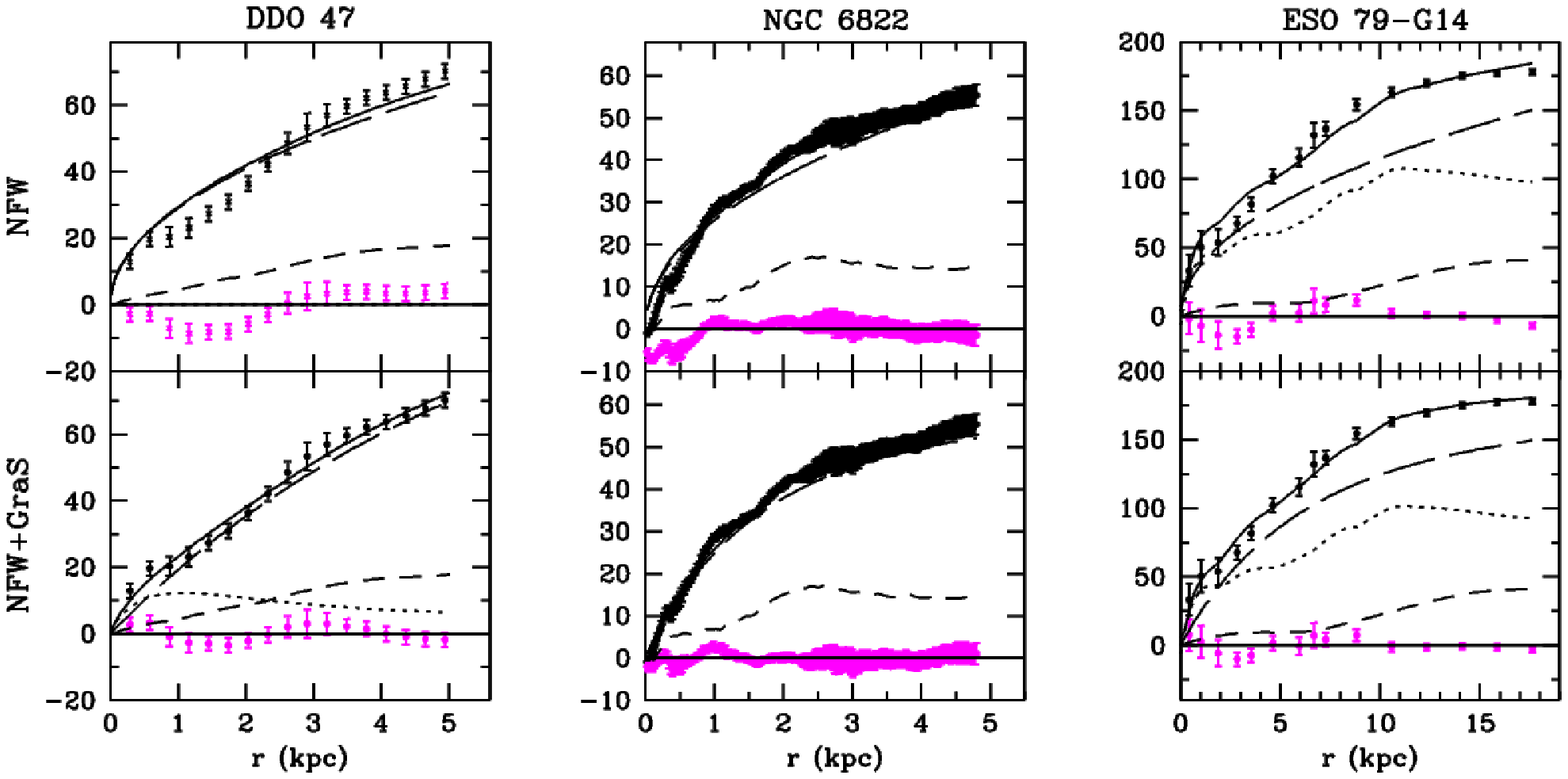}
\includegraphics[width=14.85cm]{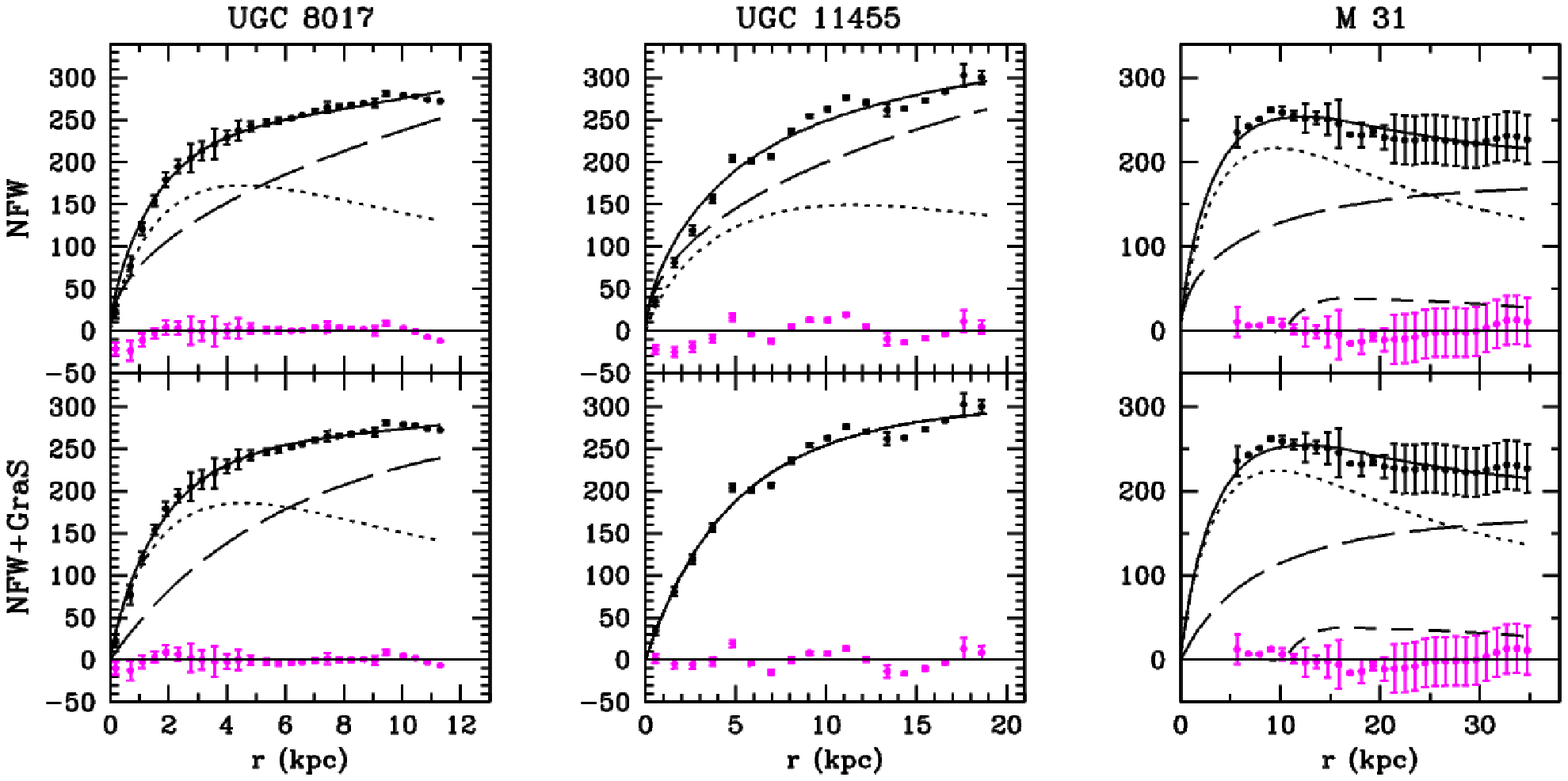}
\vskip  -0.25truecm
\caption{Galaxies in which GraS does not solve the core versus cusp discrepancy  controversy. The fitting values of the mass-to-light ratio (NGC 6822, ESO 79-G14, UGC 11455) and halo mass (DDO 47, UGC 8017) result unphysical. See Fig. $1$ and Table I for details.}
\label{fig:two}
\end{figure*}
\end{centering}
\end{widetext}

\begin{centering}
\begin{figure}[htbp]
\includegraphics[width=8.6cm]{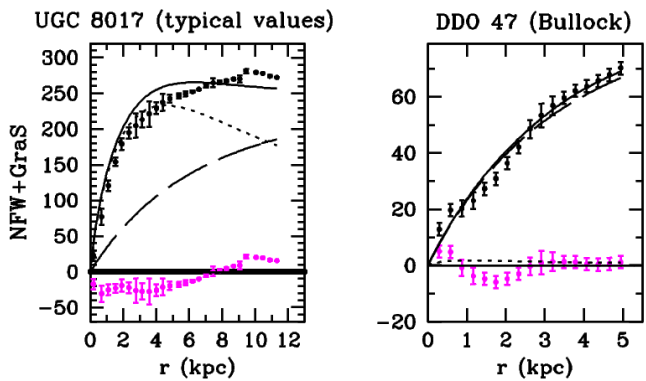}
\vskip -0.25truecm 
\caption{NFW+GraS mass model. Left: with typical values for the halo mass and the mass-to-light ratio. Right: with the Bullock \emph{et al.} relation. See Figs. 1 and 2 and Table I for details and comparison.}
\label{fig:three}
\end{figure}
\end{centering}

In detail, in the cases of ESO 116-G12 and UGC 10891, we have that GraS fits sufficiently well the RCs unlike the NFW, confirming that this model could work in some objects (see Fig. $1$, Table I).

However, in the other cases, although the fits are satisfactory, the best-fit values of the halo mass and mass-to-light ratio are unphysical.
In fact, we expect (see above) the mass-to-light ratios for NGC 6822, ESO 79-G14, UGC 11455, to be equal to (1, 2.6, 3.5), while we found much smaller best-fit values ($<$0.02, 0.3, $<$0.2).
In the same way, we expect halo masses for DDO 47 and UGC 8017 to be equal to ($9\times 10^{10}\: M_{\odot}$, $1.9\times 10^{12} \: M_{\odot}$), while we found much bigger best-fit values ($8.1\times 10^{11}\: M_{\odot}$, $1.5\times 10^{14} \: M_{\odot}$).
Furthermore, in M 31 the GraS modification is negligible and irrelevant (see Fig.\:$2$, Table I).

Let us notice that by constraining the values  for  the mass  parameters within physically acceptable values, we obtain unacceptable fits for the GraS mass model, similar to those of the Newtonian NFW case.
As an example, in UGC 8017 with $M_{halo}=3\times 10^{12}M_{\odot}$ and $M_{D}/L_{B}=\: 3 M_{\odot}/L_{\odot}$, GraS shows an unacceptable fit to data (see Fig. 3).
More in general, we realize that for all six objects, all values of $\rho_{s}$ and $r_{s}$ within their  $1\sigma$ uncertainties imply unphysical halo masses and/or mass-to-light ratios. 

We now implement the Bullock \emph{et al.} concentration vs  halo mass  relation, that eliminates one parameter in the original NFW profile.
With this relation built in, GraS performs even worse than before. See in Fig. 3 the case for DDO 47. 

In conclusion, the GraS-PM model fails to rescue the NFW profiles  in a number of  high quality well-suited RCs.
Moreover, let us point out that there is not  a pattern of  this inability, so that it is presently difficult to understand  how to  modify it in order to reach its original goal. 
Then the GraS model is a rather strong hypothesis that does not seem solve the core versus cusp discrepancy problem of the mass distribution of the center of dark matter halos.

Finally, let us remark that also in this Letter it has emerged that the available kinematics of galaxies is very constraining for non-Newtonian theories of gravity.

We warmly thank G. Veneziano, F. Piazza, P. Chimenti, I. Yegorova, G. Gentile and S. Petcov for helpful discussions. 
We appreciate the detailed comments from the anonymous referee who helped to improve the final version of this Letter.
This research was supported by CAPES-Brasil (C.F.M).

\end{document}